\documentclass[
    twocolumn,
	prd,
        amssymb,
	preprintnumbers,
	secnumarabic,
	nofootinbib,
	superscriptaddress]{revtex4-1}

\pdfoutput=1

\usepackage{graphicx}
\usepackage{enumitem}
\usepackage{latexsym}
\usepackage{amsfonts}
\usepackage{amssymb}
\usepackage{pifont}
\usepackage{color}
\usepackage{xcolor}
\usepackage{amsmath}
\usepackage{slashed}
\usepackage{dcolumn}
\usepackage{verbatim}
\usepackage{float}
\usepackage{multirow}
\usepackage{xspace}
\usepackage[normalem]{ulem}
\usepackage{hyperref}

\definecolor{hypershade}{rgb}{0.3,0.3,0.8}
\hypersetup{
  pdfauthor={Nirmal Raj},
  pdftitle={Smoke and mirrors},
  pdfsubject={Mirror neutrons},
  colorlinks=true,
  citecolor=red,
  urlcolor=blue,
  linkcolor=hypershade
}

\newcommand{\gsim}{\gtrsim}
\newcommand{\lsim}{\lesssim}

\def\Oc{\mathcal{O}}

\newcommand{\beq}{\begin{eqnarray}}
\newcommand{\eeq}{\end{eqnarray}}
\newcommand{\bea}{\begin{eqnarray}}
\newcommand{\eea}{\end{eqnarray}}
\newcommand{\nn}{\nonumber}

\def\xmark{\ding{55}}

\allowdisplaybreaks

\begin{document}

\title{Could compact stars in globular clusters constrain dark matter?}

\author{Raghuveer Garani}
\email{garani@fi.infn.it}
\affiliation{INFN Sezione di Firenze, Via G. Sansone 1, I-50019 Sesto Fiorentino, Italy}

\author{Nirmal Raj}
\email{nraj@iisc.ac.in}
\affiliation{Centre for High Energy Physics, Indian Institute of Science, C. V. Raman Avenue, Bengaluru 560012, India}

\author{Javier Reynoso-Cordova}
\email{javierisreal.reynosocordova@unito.it}
\affiliation{Dipartimento di Fisica, Universit`a di Torino, via P. Giuria 1, I–10125 Torino, Italy}
\affiliation{Istituto Nazionale di Fisica Nucleare, Sezione di Torino, via P. Giuria 1, I–10125
Torino, Italy}

\date{\today}

\begin{abstract}

The dark matter content of globular clusters, highly compact gravity-bound stellar systems, is unknown.
It is also generally unknow{\em able}, due to their mass-to-light ratios typically ranging between 1$-$3 in solar units, accommodating a dynamical mass of dark matter at best comparable to the stellar mass.  
That said, recent claims in the literature assume densities of dark matter around 1000 GeV/cm$^3$ to set constraints on its capture and annihilation in white dwarfs residing in the globular cluster M4, and to study a number of other effects of dark matter on compact stars. 
Motivated by these studies, we use measurements of stellar kinematics and luminosities in M4 to look for a dark matter component via a spherical Jeans analysis; we find no evidence for it, and set the first empirical limits on M4's dark matter distribution. 
Our density upper limits, a few $\times \ 10^4$~GeV/cm$^3$ at 1 parsec from the center of M4, do not negate the claims (nor confirm them), but do preclude the use of M4 for setting limits on non-annihilating dark matter kinetically heating white dwarfs, which require at least $10^5$~GeV/cm$^3$ densities.
The non-robust nature of globular clusters as dynamical systems, combined with evidence showing that they may originate from molecular gas clouds in the absence of dark matter, make them unsuitable as laboratories to unveil dark matter's microscopic nature in current or planned observations.

\end{abstract}

\maketitle

\section{Introduction}

Globular clusters, a.k.a. globulars, appear to surround all galaxies. 
Weighing $\Oc(10^5) M_\odot$ and spanning $\Oc$(1) pc, they are extremely dense spherical collections of stars bound by gravity. 
Unlike dwarf spheroidal galaxies (of similar masses but much greater spatial extent), no non-baryonic mass content is required to account for the stellar dynamics of globular clusters; this is in fact just what differentiates star clusters from galaxies~\cite{clustervgalaxyWillman:2012uj,clustervgalaxyTollerud}. 
This point is further corroborated by studies that fail to find compelling empirical evidence for a dark matter (DM) component in several globulars, which instead are only able to set upper bounds on the DM density; see Table~\ref{tab:gclist}.

There is no widely accepted theory of the origin of globular clusters. 
One set of models~\cite{SearleZinn:1978ApJ,*Peebles:1984ApJ,*Diemand:2005MNRAS} assumes that they are formed in DM subhalos approximately at the same time as the host galaxy. 
It then suggests that after their formation they merged with the Galactic halo, followed by severe tidal stripping, leaving the clusters with small $\Oc(1)$ mass-to-light ratios; this picture is also borne out by N-body simulations~\cite{Mashchenko:2004hj,*Moore:2005jj,*Saitoh:2005tt,*Griffen2010,*Rossi2016,*Creasey2018,*ReinaCampos2019,*Vitral:2021jvp}.  
Other models suggest that globular clusters may have formed in DM-poor environments, i.e., from giant star-forming molecular gas clouds that either collapse~\cite{Kravtsov:2003sm,*Claydon_2019} or get compressed by shock waves from galaxy mergers, as observed in the Antennae Galaxies~\cite{Ashman2001,*vandenBergh2001}. 
In any case, an important hint on globular cluster formation comes from the observation of a linear relation between the total mass of globulars and the mass of their parent halo~\cite{Hudson:2014mva,*Harris2017,*BoylanKolchin2017}, suggesting that the globular formation rate is proportional to the available initial gas mass, which in turn must be proportional to the initial halo mass.
It also suggests that globular clusters form early in the galaxy's history before star formation is suppressed by feedback mechanisms. 

In scenarios of globular cluster formation in DM-rich environments, it is estimated and suggested by simulations~\cite{Griffen2010,Madau:2019srr} that, today, in the cores of globular clusters only an $\Oc(10^{-4})-\Oc(10^{-3})$ fraction of the DM from the original subhalo is left over from tidal stripping. 
(The dispersal of stellar material, on the other hand, may be hindered by high pressures generated by gas collisions~\cite{Efremov1997}.) 
Therefore, as cradle subhalos typically weigh $10^6-10^8~M_\odot$, the DM content of globulars could weigh anywhere between $10^2 M_\odot$ and $10^5 M_\odot$.
For a scale radius $r_s$ of 5 pc, the scale density $\sim$ (cluster mass)/$r_s^3$ could thus range from a few GeV/cm$^3$ to upwards of $10^3$~GeV/cm$^3$.
This, indeed, is the range spanned by estimates quoted in the literature of the core density of one of the most studied nearby globulars, M4 (NGC 6121); e.g., Ref.~\cite{Hooper:2010es} near the lower end, and Ref.~\cite{McCullough:2010ai} near the upper end\footnote{Both estimates account for a modest enhancement in DM density as DM orbits contract in response to baryons collecting closely.}. 

If the estimate of Ref.~\cite{McCullough:2010ai} were true, it would have dramatic implications for models of particle DM. 
Firstly, high DM densities in globular clusters would benefit searches that look for DM annihilation products~\cite{Zaharija2008,Abramowski_2011,Feng_2012,Hurst:2014uda,Brown:2019whs,Wirth_2020,ReynosoCordova2021,Hooper:2016rap}. 
More pertinently, as pointed out in Ref.~\cite{McCullough:2010ai}, DM particles in M4/NGC6121 could capture in its white dwarf (WD) population via scattering on nuclei, 
self-annihilate to Standard Model (SM) states in their interior, and overheat the stars. 
The observed luminosities of the WDs in M4 would then place wide-ranging constraints on DM-nucleon cross sections, often outdoing underground direct detection searches.
Motivated by this, a number of papers have since appeared exploring far-reaching particle physics implications of globular cluster DM capturing in celestial bodies~\cite{Kouvaris:2010jy,Cermeno_2018,Dasgupta2019,Panotopoulos_2020,Leane:2021ihh,Bell:2021fye,Biswas:2022cyh,Ramirez-Quezada:2022uou}, all of them assuming a DM density around 1000 GeV/cm$^3$ as quoted in Ref.~\cite{McCullough:2010ai}.

In this work we quantify the DM content in M4/NGC6121 from stellar data. 
By performing a spherical Jeans analysis using measurements of stellar line-of-sight velocities and surface luminosities via a Markov chain Monte Carlo (MCMC) approach, we find no evidence for DM in M4/NGC6121, and set upper bounds on its DM distribution. 
These limits are relatively weak, i.e., comparable to the visible stellar mass, mainly as a result of the M4/NGC6121 stellar kinematic data accommodating mass-to-light ratios (in solar units) $\Upsilon \sim$ 1$-$2.
This is consistent with Ref.~\cite{baumgardthilker2019} which found $\Upsilon = 1.7 \pm 0.1 \ M_\odot/L_\odot$ using N-body simulation-based fits without accounting for DM.
(Similarly, the derived V-band mass-to-light ratios of the closest ($<$ 5 kpc from the Sun) globulars are typically about 2 ${\rm M_\odot/L_\odot}$~\cite{baumgart:gaia:2021}, and consistent with theoretical expectations for stellar systems that have evolved without DM.)
Such weak bounds on DM densities in globular clusters are not an exception but the rule. 
As argued in Ref.~\cite{clustervgalaxyWillman:2012uj}, in smallish, dense stellar systems such as globular clusters and ultra-compact dwarf galaxies, the typical dynamical mass-to-light ratio of $\sim$ 1$-$5 $M_\odot/L_\odot$ makes it very difficult to determine the presence of DM from stellar kinematics even if these systems do reside in a DM halo.
This is in contrast to, say, dwarf spheroidals, which exhibit $\Upsilon \sim 10 - 100 M_\odot/L_\odot$, and are definitely known to contain DM. 
Crucially for us, as a result of this inevitable uncertainty in the DM content of globular clusters, it is well-nigh impossible to make empirical statements about whether DM can impact compact stars in them. 
That is the main message of our study.

Given this limitation, we then ask another question: could observed WDs in M4/NGC6121 possibly constrain dark {\em kinetic} heating of WDs through DM capture?  
This is a possibility for DM with self-annihilation cross sections that are negligible or, as in the case of asymmetric DM, perhaps zero; the WD heating comes entirely from the transfer of DM kinetic energy during capture.
As the kinetic energy of DM falling into WDs is at most $\sim 10^{-2} \times$ the mass energy, much higher DM densities are required for this heating process to be interesting. 
We find that the upper bounds we have obtained on the DM densities in M4 {\em are} tight to disfavour the usefulness of this mechanism over a wide range of parameters. 
Moreover, our results also impact other scenarios in the literature where DM densities much higher than in the solar neighborhood were assumed in globular clusters, involving capture in neutron stars (NSs), triggering thermonuclear explosions in WDs, and DM in the form of primordial black holes (PBHs).
Our paper comments on them.

It is organized as follows.
In Section~\ref{sec:limits} we derive limits on the DM content of M4/NGC6121 using spherical Jeans and MCMC analyses.
In Section~\ref{sec:DMvWD} we discuss the implications of our limits for DM heating of WDs and other scenarios of DM confronting compact stars in globular clusters.  
In Section~\ref{sec:discs} we provide a summary and the scope of our work. Appendices~\ref{app:MCMCpipeline} and~\ref{app:WDlumitemp} provide technical details, and Appendix~\ref{app:otherglobclusts} surveys efforts to look for DM in globular clusters and luminosity measurements of WDs in them.

\renewcommand{\arraystretch}{1.5}
\begin{table*}[t]
  \centering
  \begin{tabular}{|l|c|c|l|l|l|}
  \hline
  globular & $d$ (kpc)  & $M/ L_{\rm V}$  ($\odot$) & DM hint? & WDs ? \\ \hline
    \hline
         NGC 6121 (M4)  &  2.2 & 1.7 $\pm$ 0.1 & \xmark \ [this work] & \cite{Bedin:2009it}  \\     \hline
        Kron 3 & 61.0 & 1.2 $\pm$  0.3 & \xmark~\cite{Lane0910} & \\ 
         NGC 121 & 64.9 & 0.9 $\pm$ 0.3 & \xmark~\cite{Lane0910} & \\
          NGC 1851  & 12.0 & $1.3 \pm 0.2 $  &\xmark~\cite{ReynosoCordova2021} & \\
     NGC 2808  & 10.0 & $1.4 \pm 0.1 $  &\xmark~\cite{Baumgardt2009,ReynosoCordova2021} & \cite{WDsNGC2808Dieball:2005xh,Moehler:2008ai} \\
      NGC 3201  & 4.7 & $2.6 \pm 0.1 $  &\xmark~\cite{ReynosoCordova2021} & \\
     NGC 4590 (M68)  & 10.4  & $1.9 \pm 0.1 $  &\xmark~\cite{Lane0908} &  \\
   NGC 5024 (M53)  & 18.5  & $2.0 \pm 0.1 $ &\xmark~\cite{Lane0908} &  \\
     NGC 6093 (M80)  & 10.3  & $2.0 \pm 0.1 $  &\xmark~\cite{Baumgardt2009,ReynosoCordova2021} & \cite{Moehler:2008ai} \\
     NGC 6656 (M22)  & 3.3  & $2.0 \pm 0.1 $  & \xmark~\cite{peterson:1986,Lane0908,Reynoso-Cordova:2022ojo} & \cite{Piotto:1999em,WDsM13M22}\\
     NGC 6752  & 4.1  & $2.2 \pm 0.1 $ &\xmark~\cite{Baumgardt2009,Reynoso-Cordova:2022ojo}  &  \cite{WDsNGC6752Chen2022,bedin6752:2023}\\
     NGC 6397  & 2.3 & 2.4 $\pm$ 0.5 & \xmark~\cite{Heyl:6397:2012ApJ,eduardo:2021,pierre:2021}  & \cite{Richer:2007cv,Hansen:2007ve,WDsNGC6397Torres2015}\footnote{contains the coldest WD (surface $T \simeq 3700 K$) observed~\cite{Hurst:2014uda}.}  \\
         NGC 6809 (M55)  & 5.4 &  2.1 $\pm$ 0.1&\xmark~\cite{Lane0910} & \\ 
     NGC 6838 (M71)  & 4.0 & 1.0 $\pm$ 0.05 &\xmark~\cite{richerm71:1989,1996:heggie} & \cite{richerm71:1989} \\ 
      NGC 7078  & 10.7  & $1.3 \pm 0.1$ & \xmark~\cite{Baumgardt2009} & \cite{Moehler:2008ai} \\
      NGC 7089 (M2)  & 11.7  & $1.8 \pm 0.1$ & \xmark~\cite{ReynosoCordova2021} &  \\ 
     NGC 7099 (M30)  & 8.5  & $1.6 \pm 0.1$ & \xmark~\cite{Lane0908,ReynosoCordova2021} &  \\ \hline
      NGC 104 (47 Tuc)  & 4.5  & $1.9 \pm 0.1 $ &  \xmark~\cite{1996:heggie,Lane0910,Reynoso-Cordova:2022ojo} &  \cite{enrique:2014,goldsbury:2016,WDs47Tuc}\\
    &   &   & \checkmark~\cite{Brown:2018pwq} & \\ 
    NGC 2419 & 88.5 & 1.6$\pm$0.2 & \xmark~~\cite{Baumgardt2009,Conroy2011,Ibata2012} & \\ 
    & & & \checkmark~\cite{Ibata2012} & \\
      NGC 3201  & 4.7  & $2.6 \pm 0.1 $ &  \xmark~\cite{wan:2021} & \cite{Vitral:2021jvp,Vitral:2022apu}\\
    &   &   & \checkmark~\cite{bianchini:2019} & \\ 
        NGC 5139 ($\omega$ Cen) & 5.4 & $2.8 \pm 0.1$  &  \xmark~\cite{Reynoso-Cordova:2022ojo} & \cite{WDsomegaCenCalamida:2007hr,Cool:2012cb} \\
    &   &   & \checkmark~\cite{Brown:2019whs,EvansStrigari:2021bsh} &  \\ 
      NGC 6544  & 2.5  & $2.3 \pm 0.5 $  &?~\cite{minniti:2021} & \cite{Lynch:2012,contreras:2017}\\ \hline
      NGC 5128 population  & 3--5 Mpc  & $>6$~\cite{Taylor2015darkglobs}  &?~\cite{Taylor2015darkglobs} & \\ \hline 
 \end{tabular}
  \caption{A non-exhaustive list of globular clusters in which a dark matter component was searched for using stellar data.  
  Except for the globular population in the elliptical galaxy NGC 5128, the distances from the Sun $d$ and mass-to-light ratios in the V-band in units of $M_\odot/L_\odot$ are taken from Ref.~\cite{baumgart:gaia:2021}. 
  A `\xmark' indicates DM was not found by the study cited, a `$\checkmark$' indicates statistical evidence, and `?' indicates ambiguous conclusions.
  The last column lists references on observations of white dwarfs, which in principle may be  used to set limits on DM-induced heating if unambiguous evidence for a DM component is found in the corresponding globular cluster.}
  \label{tab:gclist}
\end{table*}

\section{Limits on dark matter density in M4/NGC6121}
\label{sec:limits}

\begin{figure*}
    \centering
    \includegraphics[width=0.45\textwidth]{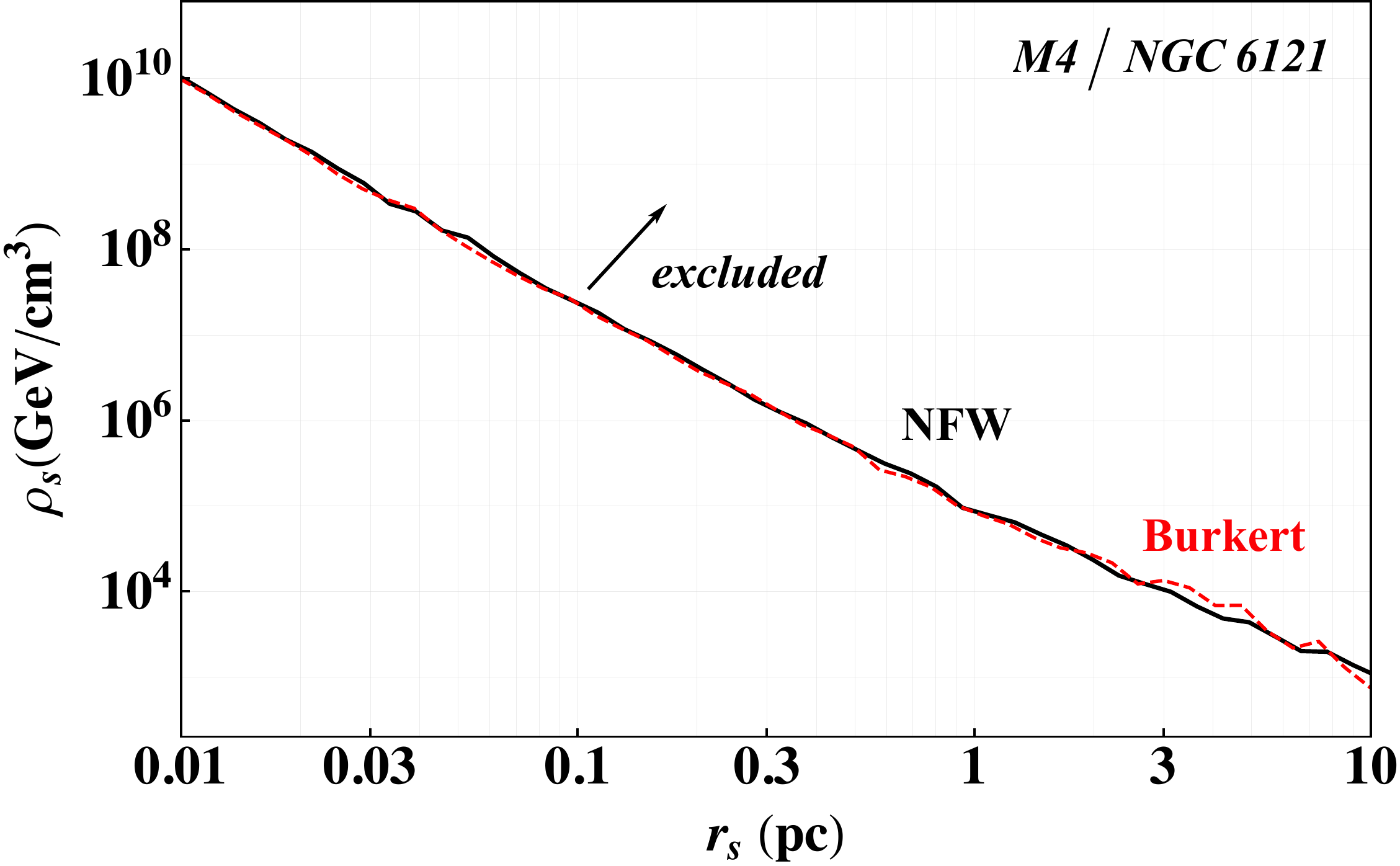}
    \includegraphics[width=0.46\textwidth]{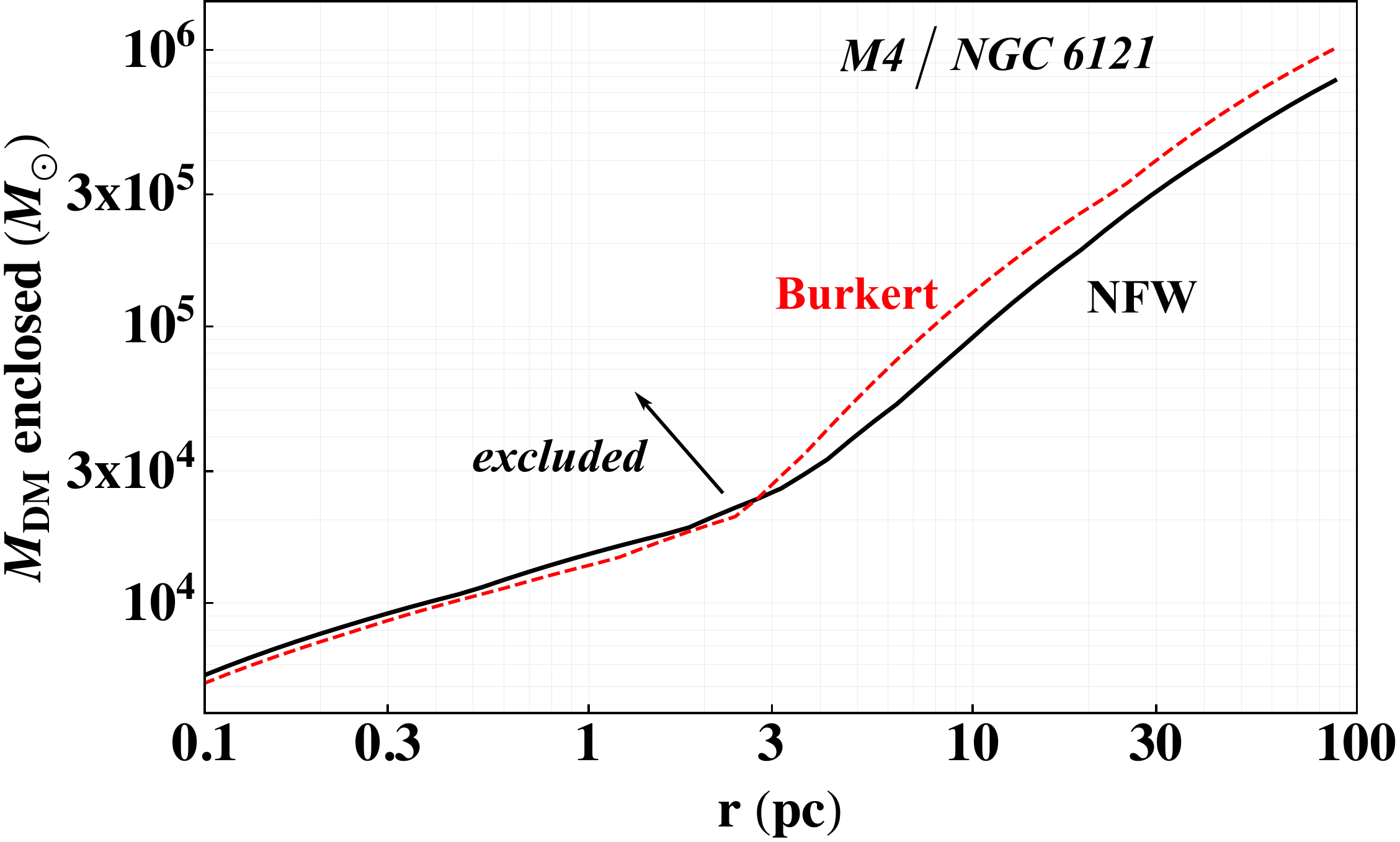} \\
\hspace{-.3cm} \includegraphics[width=0.445\textwidth]{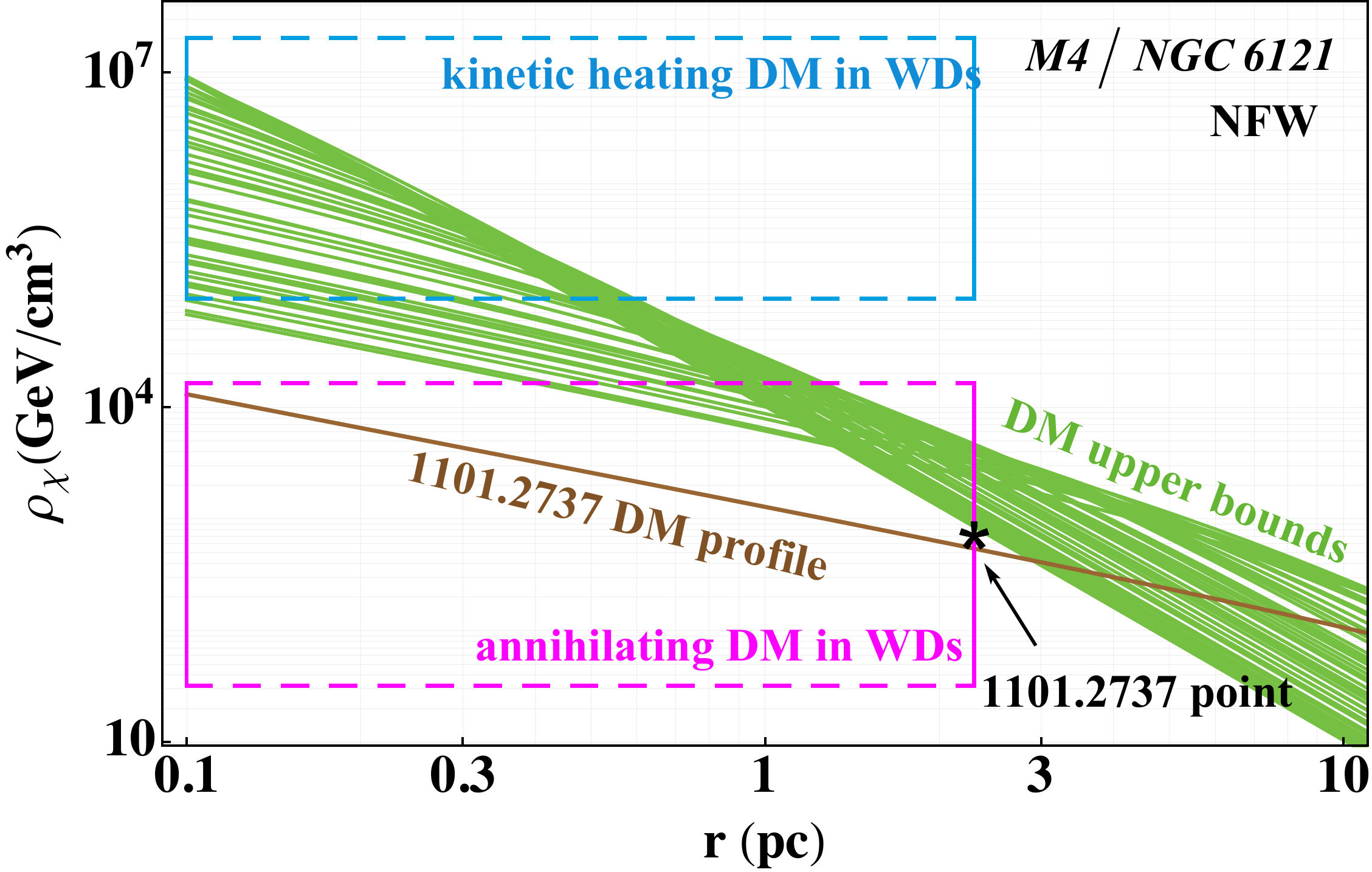} 
  \includegraphics[width=0.445\textwidth]{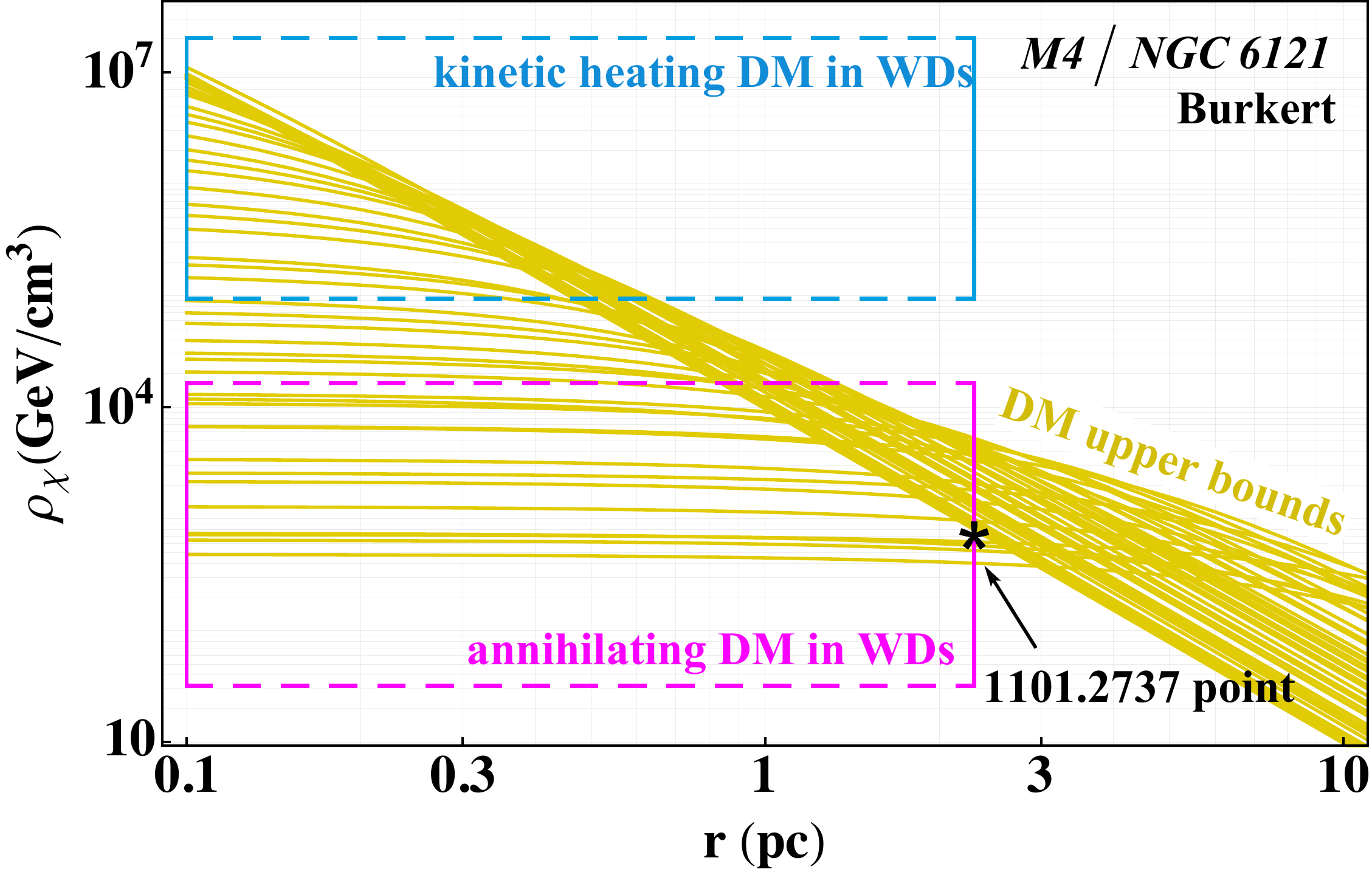}
    \caption{{\bf \em Top left}: 95\% C.L. upper limits on the scale density vs scale radius of an assumed NFW or Burkert profile of dark matter in the M4/NGC6121 globular cluster. 
    {\bf \em Top right}: 95\% C.L. upper limits on the enclosed halo mass.
    {\bf \em Bottom left}: A collection of NFW DM density profiles corresponding to $\{r_s,\rho_s\}$ in the top left panel.  
    The magenta lines enclose regions with observed WD luminosities (and inferred radii \& masses) converted to an equivalent DM density using the DM capture rate, assuming WDs are heated by DM self-annihilations within.
    The cyan lines are the same, but assuming WDs are heated by transfer of DM kinetic energy alone.
    The horizontal span of these lines reflect the uncertainty in WD positions, and their vertical span reflect the range of WD luminosities.
    The brown curve depicts the NFW profile for parameters estimated using a spherical collapse model in Ref.~\cite{McCullough:2010ai};
 the asterisk denotes the DM density at $r = 2.3$~pc in this model after accounting for adiabatic contraction.
 {\bf \em Bottom right}: Same as bottom left, but for the Burkert DM density profile. 
 See text for further details.
    }
    \label{fig:densitylimitsM4NFW}
\end{figure*}

The top left panel of Figure~\ref{fig:densitylimitsM4NFW} shows our 95\% C.L. upper limits on the dark matter distribution in M4/NGC6121 
We display these in the plane of the scale density $\rho_s$ vs the scale radius $r_s$ of a Navarro-Frenk-White (NFW) \cite{navarro_1996} profile of DM density,
\begin{equation}
    \rho_{\rm{NFW}}(r)=\frac{\rho_s}{\bigg[\frac{r}{r_s} \bigg] \bigg[1 + \frac{r}{r_s}\bigg]^2}~,
    \label{eq:NFW}
\end{equation}
as well as a Burkert profile that is more cored in the inner halo regions,
\begin{equation}
    \rho_{\rm{Bur}}(r)=\frac{\rho_s}{\bigg[ 1+ \frac{r}{r_s}\bigg]\bigg[ 1 + \left( \frac{r}{r_s}\right)^2\bigg]}~.
    \label{eq:Burk}
\end{equation}
As we have not found evidence for DM, and as this null result does not depend crucially on the choice of DM profile for reasons argued in the Introduction, we do not consider other profiles. 

Our limits were obtained by performing a Jeans analysis via a Markov chain Monte Carlo (MCMC) technique.
We briefly outline the method below and in detail in Appendix~\ref{app:MCMCpipeline}, but first one can gain a rough understanding of our limits as follows.

The total mass of M4/NGC6121 $\simeq 10^5 M_\odot$ and its mass-to-light ratio (which we independently determine in our fit) in solar units is about unity~\cite{baumgardt2023}. 
Thus we can expect the maximum allowed DM mass to vary between $\sim 10^4-10^6~M_\odot$. 
The mass of NFW and Burkert halos obtained by integrating over Eq.~\eqref{eq:NFW} and Eq.~\eqref{eq:Burk} is:  
\bea
\nn    M_{\rm NFW}&=& 4\pi \rho_s r_s^3 \bigg[\log(\kappa+1) - \frac{\kappa}{\kappa+1} \bigg]~,\\
M_{\rm Burk} &=& \pi \rho_s r_s^3 [\log((\kappa^2+1)(\kappa+1)^2) - 2 \tan^{-1}\kappa],\ \ 
    \label{eq:MhaloNFW}
\eea
where $\kappa \equiv r_{\rm max}/r_s$ determines the radius $r_{\rm max}$ at which the halo is truncated;
the $\kappa$-dependent term in Eq.~\eqref{eq:MhaloNFW} is an $\Oc(1)$ number. 
It may now be seen that the exclusion \{$r_s, \rho_s$\} in Fig.~\ref{fig:densitylimitsM4NFW} do indeed give $M_{\rm NFW}$ and $M_{\rm Burk}$ in the maximum allowed range.

Of course, it is not the total mass, but the mass {\em profiles} (both dark and stellar) of a structure that determine its velocity dispersion profile.
To estimate these profiles we perform a spherical Jeans analysis~\cite{rounddenim} assuming that M4/NGC6121 is virialized.
This is a justified assumption since its half-mass relaxation time, about 0.87 Gyr~\cite{baumgardt2023}, is much shorter than its age inferred from the cooling sequence of its constituent WDs, about 12 Gyr~\cite{Bedin:2009it}. 
 Assuming equal polar and azimuthal dispersion speeds, $\sigma_\theta^2=\sigma_\phi^2$, the stellar population can be described by the non-collisional Jeans equation: 
\begin{equation}
\rho^{-1}_\star(r)\frac{\partial}{\partial r} \left( \rho_\star(r) \sigma_r^2 \right) + \frac{2 \beta(r) \sigma_r^2}{r} = - \frac{G M_{\rm enc}(r)}{r^2},
\label{eq:boltzmann}
\end{equation}
where $\sigma_r$ is the radial dispersion velocity, 
$\beta \equiv 1 -\sigma_\theta^2/\sigma^2_r$ is the anisotropy, 
$\rho_\star(r)$ is the stellar density profile,
and $M_{\rm enc}(r) = M_\star (r) + M_{\rm DM} (r)$ is the total mass enclosed within a radius $r$, with $M_\star (M_{\rm DM})$ the stellar (DM) mass.

For our fit we use stellar line-of-sight (LOS) velocity data from the Very Large Telescope and Keck Observatory as collected in Ref.~\cite{2018MNRAS.478.1520B}. 
Solving Eq.~\eqref{eq:boltzmann} for the radial velocity dispersion, we project it on to the LOS:
\begin{equation}
\sigma_{\rm{LOS}}(r)=\frac{2}{\Sigma_\star(r)} \int_{r}^\infty dr' \bigg( 1 - \beta(r') \frac{r^2}{r'^2}\bigg) \frac{r'\rho_\star(r')\sigma^2_r(r')}{\sqrt{r'^2 -r^2}},
\label{eq:sigma_los}
\end{equation}
where $r'$ is the 2D-projected radius and $\Sigma_\star$ the stellar tracer surface mass density.
Data on the latter is obtained from Ref.~\cite{2019MNRAS.485.4906D}, which compiles Gaia DR2 and Hubble Space Telescope surface brightness measurements for a large sample of globular clusters. 

Using the aforementioned data, and following  the procedure of Ref.~\cite{Reynoso-Cordova:2022ojo} to select members and remove possible binaries, we perform our MCMC analysis with 13 free parameters:
\bea
\nn {\rm DM \ distribution}:&& \ \{\rho_s, r_s\} \\
{\rm stellar \ distribution}:&& \ \{ \rho_\star \ {\rm parameters} \ (6), M_\star \}\\
\nn {\rm velocity \ anisotropy}:&& \ \{ \beta \ {\rm parameters} \ (4) \}~.
\label{eq:freeparams}
\eea

The stellar distribution is modeled as a sum of three Plummer spheres~\cite{Plummer1911}, and the velocity anisotropy profile is modeled as a smoothly varying two-part function; we describe these in detail in Appendix~\ref{app:MCMCpipeline}.  
Due to tidal stripping, the stellar profile is in principle truncated at some radius $r_t$, which is not accommodated by a Plummer sphere decomposition.
However, in practice, the data points available for surface luminosity and velocity dispersion are at radii far below $r_t \simeq$ 50 pc as reported in Ref.~\cite{baumgardt2023} or $r_t \simeq$ 20 pc as estimated in Ref.~\cite{McCullough:2010ai}, and thus the effect is negligible.
It is also reasonable to assume that the DM profile is truncated at $r = r_t$, and again the effect of not including this truncation in practice is negligible.

 \begin{figure}[htb!]
    \centering
    \includegraphics[width=0.45\textwidth]{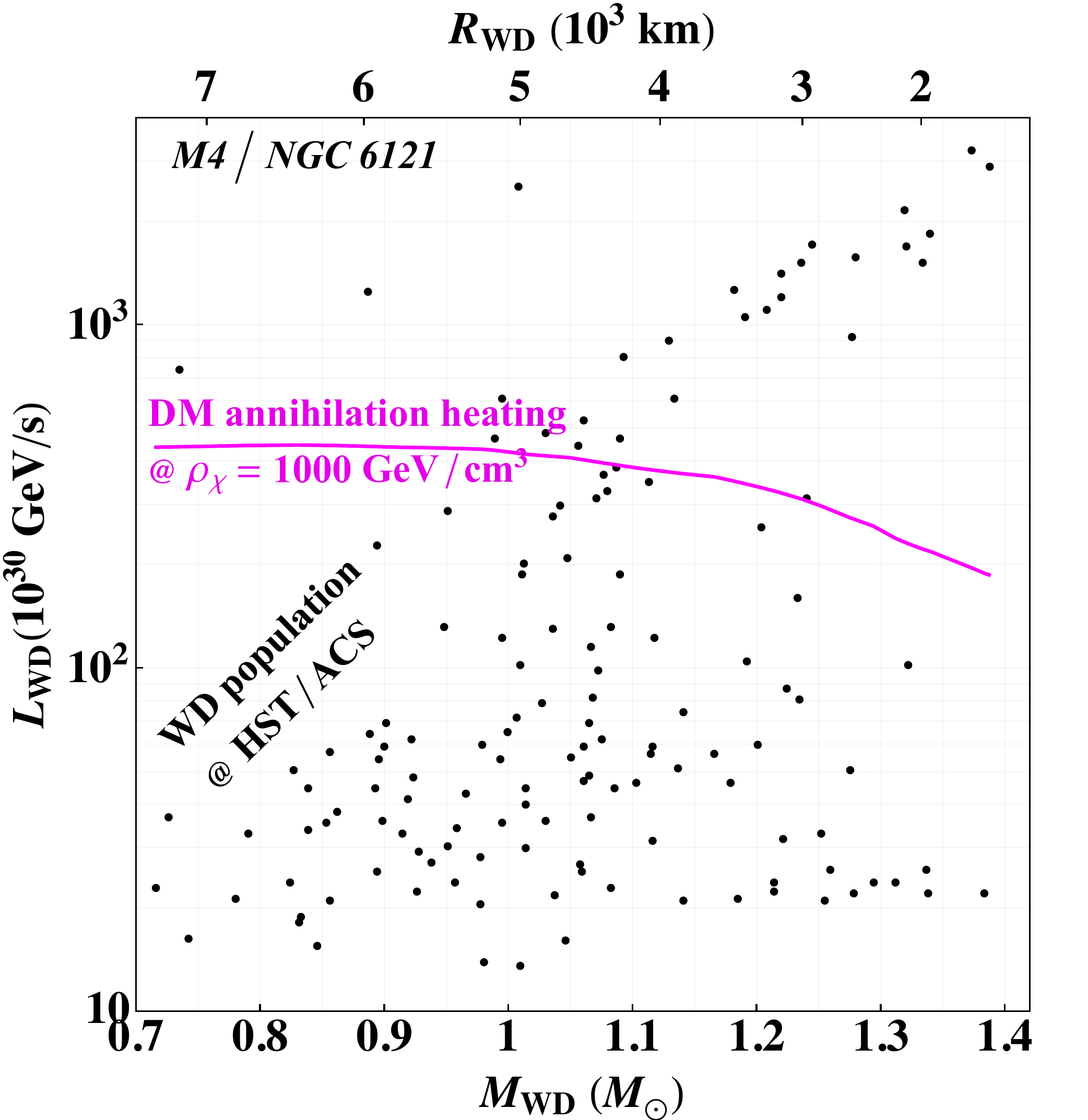}
    \caption{
  The white dwarf population of M4/NGC6121 observed in HST/ACS in the luminosity-mass plane obtained from the color-magnitude diagram in Ref.~\cite{Bedin:2009it} using the routine described in Appendix~\ref{app:WDlumitemp}.
  The WD radius ticks in the top x-axis are obtained from a WD mass-radius relation via a Feynman-Metropolis-Teller equation of state.
  Also shown is a curve of the WD luminosity imparted by dark matter heating via annihilations within the WDs, assuming a DM density of 1000 GeV/cm$^3$.
  The WD points below this curve would lead to constraints on DM capture.
  An analogous curve, ranging around $L_{\rm WD} = 10^{29}$~GeV/s, exists for WD heating through DM kinetic energy transfer alone, but for the sake of visual clarity we haven't displayed it.
  }
    \label{fig:LvMscatter}
\end{figure}

In the top right panel of Fig.~\ref{fig:densitylimitsM4NFW} we plot the 95\% C.L. upper limits on the DM mass enclosed within a radius $r$.
We see that the slope of these limits is steeper at large $r$ and gentler at small $r$. 
As argued in Ref.~\cite{ReynosoCordova2021} for bounds on other globulars, this is because of an observational bias in the stellar kinematic data, which is only available for $r \gsim \Oc(1)$~pc. 
For large $r$ that includes much of the kinematic data, the NFW and Burkert profiles that maximize the enclosed $M_{\rm DM}$ are those with $r_s \gsim r$.
It may seen from Eq.~\eqref{eq:MhaloNFW} that for small $\kappa$ (where now $r_{\rm max} \to r$) the enclosed $M_{\rm DM} \propto r_s r^2$ for NFW and $M_{\rm DM} \propto r^3$ for Burkert profiles -- hence the steeper Burkert bound in the plot.
On the other hand, for $r \lsim \Oc(1)$~pc, the profiles that maximize the enclosed $M_{\rm DM}$ are those with $r_s \lesssim O(1)$~pc: 
now the first term in Eq.~\eqref{eq:MhaloNFW} dominates and the enclosed $M_{\rm DM} \propto r_s^3 \log(r/r_s)$ for both profiles.
  
In the next section we will discuss the implications of these results for DM interactions with compact stars.

\section{Dark matter, compact stars, and globular clusters}
\label{sec:DMvWD}

We now turn to the question of whether the allowed DM densities rule out the use of compact stars in globular clusters as detectors of DM.
We discuss first white dwarfs as thermal detectors, then other signatures of dark matter encountering compact stars in general.

\subsection{Dark matter capture and heating of white dwarfs}

 DM particles intercepted by compact objects can be efficiently captured in their deep gravitational potential wells by losing energy via scattering on stellar constituents; see Refs.~\cite{snowmassWP:Berti:2022rwn,Goldman:1989nd,*Bramante:2017xlb,*NSvIR:Baryakhtar:DKHNS,*NSvIR:Raj:DKHNSOps,*NSvIR:GaraniGenoliniHambye,*NSvIR:GaraniHeeck:Muophilic,*NSvIR:Pasta,*NSvIR:Riverside:LeptophilicShort,*NSvIR:Riverside:LeptophilicLong,*NSvIR:DasguptaGuptaRay:LightMed,*NSvIR:clumps2021,*NSvIR:IISc2022,*NSvIR:pseudoscalarmodelsMF} and references therein.
 Assuming a Maxwell-Boltzmann distribution of DM velocities with dispersion $v_d$, the rate of DM capture in a WD of mass $M_{\rm WD}$ and radius $R_{\rm WD}$ is given by~\cite{Tinyakov:2021lnt}
 \begin{equation}
     C_{\rm \chi} =  \frac{\rho_\chi}{m_\chi} \pi R_{\rm WD}^2 \, \frac{\gamma^2-1}{v_\star}  \,{\rm erf}\left(\sqrt{\frac{3}{2}} \frac{v_\star}{v_d}\right) \times p_\sigma~,
     \label{eq:caprate}
 \end{equation}
 in the limit where the WD surface escape speed $v_{\rm esc}=\sqrt{2GM_{\rm WD}/R_{\rm WD}} \gg v_d, v_\star$, with $v_\star$ the WD speed.
 Here $\gamma = (1-v^2_{\rm esc})^{-1/2}$ and $p_\sigma = 1-e^{-\tau}$ is the probability for incident DM to scatter given an optical depth $\tau$.
 In the optically thin limit, $p_\sigma \simeq \tau = \sigma_{\chi T}/\sigma_{\rm geo}$, where $\sigma_{\chi T}$ is the DM cross section for scattering on target $T$ (nucleus or electron), and $\sigma_{\rm geo}$ is the WD geometric cross section, $\sigma_{\rm geo} = \pi R_{\rm WD}^2/N_T$, where $N_T$ is the number of targets in the WD.  
 For simplicity we assume below that WDs are composed dominantly of $^{12}$C.

 DM capture adds energy to the WD medium by transfer of kinetic energy at a rate $\dot Q_{\rm kin} = (\gamma-1)m_\chi C_{\rm \chi}$.
 In some DM scenarios, captured DM possibly self-annihilates to SM states and heats the WD further at a rate $\dot Q_{\rm kin+ann} = \gamma m_\chi C_{\rm \chi}$.
 As $\gamma \simeq 1$ for WDs it is the latter mechanism, if available, that dominates WD heating.
 Under thermal equilibrium the WD luminosity equals the DM heating rate, and
 for DM-nucleus scattering cross section equal to or below the geometric value and DM mass above the WD evaporation mass $\sim \Oc({\rm MeV})$~\cite{Garani:2021feo,Bell:2021fye}, we obtain a blackbody temperature (as seen by a distant observer) of
 \begin{eqnarray}\label{eq:teffk}
T_{\rm kin}^{\infty} &\approx&  1100\,\rm{K}\Bigg[\frac{\alpha_{\rm kin}}{3\times 10^{-7}} \left(\frac{\rho_\chi}{10^3 \,\rm{GeV/cm^3}}\right) \left(\frac{\sigma_{\chi N}}{\sigma_{\rm geo}} \right) \nonumber \Bigg. \\
&& \hspace{-1cm}\Bigg. \times \left(\frac{20\,\rm{km/s}}{v_\star}\right) {\rm{erf}}\left( \sqrt{\frac{3}{2}} \frac{10\,\rm{km/s}}{v_d} \frac{v_\star}{20\,\rm{km/s}} \right) \Bigg]^{1/4}\, , 
\end{eqnarray}
\begin{eqnarray}\label{eq:teffka}
T_{\rm kin+ann}^{\infty} &\approx& 7700~\rm{K}\Bigg[ \frac{\alpha_{\rm kin+ann}}{8\times 10^{-4}} \left(\frac{\rho_\chi}{10^3 \,\rm{GeV/cm^3}}\right)  \left(\frac{\sigma_{\chi N}}{\sigma_{\rm geo}} \right)  \nonumber \Bigg. \\ 
& & \hspace{-1cm}\times \Bigg. 
\left(\frac{20\,\rm{km/s}}{v_\star}\right) {\rm{erf}}\left(\sqrt{\frac{3}{2}} \frac{10\,\rm{km/s}}{v_d} \frac{v_\star}{20\,\rm{km/s}} \right)  \Bigg]^{1/4} \, ,
\end{eqnarray}
where
\beq\label{eq:alpha}
\alpha_{\rm kin} =  \frac{(\gamma -1)(\gamma^2-1)}{\gamma^4} \quad \quad \rm{and} \quad \quad \alpha_{\rm kin+ann} = \frac{\gamma (\gamma^2-1)}{\gamma^4}~.\nonumber
\eeq
In the above equations we have normalized quantities to values corresponding to $M_{\rm WD} = 1.2 \,M_\odot$ and $R_{\rm WD} = 4000$ km, and to average dispersion speeds in globular clusters~\cite{HarrisGCcat:1996AJ,baumgart:gaia:2021}.
DM models can in principle be constrained by
observing WDs colder than Eqs.~\eqref{eq:teffk} and \eqref{eq:teffka}.
In Fig.~\ref{fig:LvMscatter} we show a WD population observed in M4/NGC6121 in the plane of luminosities and masses;
we derived this data from HST-ACS observations~\cite{Bedin:2009it} using a procedure described in detail in Appendix~\ref{app:WDlumitemp}.
We also show here a contour corresponding to the maximal WD heating via DM annihilations (i.e. $p_\sigma = 1$ in Eq.~\eqref{eq:caprate}) assuming $\rho_\chi$ = 1000 GeV/cm$^3$.
Clearly, for this value of DM density several WDs are fainter than they would be in the presence of DM annihilations and may be used to set limits on DM capture.

Information on a WD's mass, radius and luminosity $L_{\rm WD}$ can be used to determine the minimum ambient DM density $\rho^{\rm WD}_{\rm \chi, min}$ required to constrain its heating the WD by requiring $\dot Q \geq L_{\rm WD}$.
For the WD population in Fig.~\ref{fig:LvMscatter} the span of $\rho^{\rm WD}_{\rm \chi, min}$ is shown in the bottom left panel of Fig.~\ref{fig:densitylimitsM4NFW} with the magenta (cyan) region corresponding to heating from DM annihilations (kinetic energy transfer). 
The horizontal span of these regions denotes the uncertainty in the cluster-centric distance $r$ of the WDs. 
Specifically, the cutoff at $r = r_{\rm max}$ = 2.3 pc corresponds to the maximum angular distance of 250$^{\prime \prime}$ at which WDs were observed at HST/ACS, and the cutoff at $r = r_{\rm min}$ = 0.1 pc corresponds to an estimate of the minimum distance at which WDs could be resolved.
In more detail, the angular resolution of  HST/ACS is about 0.1$^{\prime \prime}$~\cite{acshandbook} corresponding to $r \simeq 10^{-3}$~pc, but for crowded stellar fields in the inner regions of globular clusters it is still challenging to resolve individual stars; on the other hand, the first point at which stellar line-of-sight velocity information is available is at $r \simeq$~pc.  
We have chosen $r_{\rm min}$ = 0.1 pc as a compromise between these two.

These values of $\rho^{\rm WD}_{\rm \chi, min}$ may be compared with the green curves, which depict a span of DM NFW profiles corresponding to the upper limits on $\{r_s, \rho_s\}$ in the top left panel.
For annihilation heating, values of $\rho^{\rm WD}_{\rm \chi, min} \lsim$ 800 GeV/cm$^3$ always remain below our DM profile limits, thus previous limits on DM capture may still be valid.
In particular, the point marked with a star, depicting $\rho_\chi$ = 798 GeV/cm$^3$, is the estimation of the DM density at $r = 2.3$~pc after numerically accounting for adiabatic contraction in the subhalo collapse model of Ref.~\cite{McCullough:2010ai}.
This estimate was used to rule out DM-induced WD heating.
Our DM density upper bounds are not strong enough to invalidate this claim.
For reference, we also plot the DM density profile (in brown) for the NFW parameters estimated by Ref.~\cite{McCullough:2010ai} for the uncontracted halo.
As we had explained before, it is far from obvious that the DM density upper limits would improve in the future by orders of magnitude -- even with more precise data on stellar motion -- as $O(1) M_\odot/L_\odot$ mass-to-light ratios imply that the limit on the total DM mass cannot be much smaller than the total stellar mass.
Our conclusion here is that, due to this lack of robustness, M4/NGC6121 is an unsuitable system to constrain DM annihilation heating of WDs.

We also see that for purely kinetic heating to be relevant much higher DM densities are required to compensate for the smaller fraction of energy transferred than in DM annihilations (see Eq.~\eqref{eq:alpha}).
In fact, there are regions where all values of $\rho^{\rm WD}_{\rm \chi, min}$ for kinetic heating lie above our DM density limits.
There are other regions where our limits overlap with, or exceed, $\rho^{\rm WD}_{\rm \chi, min}$.
As before, our conclusion here is that our limits make M4/NGC6121 an unsuitable system to constrain DM kinetic heating of WDs.

These conclusions are qualitatively unchanged for a more cored DM profile, such
as a Burkert profile~\cite{Burkert_1995} that is shallower than NFW in the halo's inner regions (Eq.~\eqref{eq:Burk}).
As seen in the top panels of Fig.~\ref{fig:densitylimitsM4NFW} our results are very similar to the NFW case. 
This is due to quantitatively similar support to our fit in the inner regions of M4/NGC6121, where stellar kinematic data is poor.

The similarity of our conlusions can also be seen in the bottom right panel of Fig.~\ref{fig:densitylimitsM4NFW}, where we have drawn -- similar to the bottom left panel for NFW -- the Burkert DM density profiles corresponding to the $\{r_s,\rho_s\}$ upper limits.

\subsection{Other signatures of dark matter in compact stars}

The presence of high densities of DM in globular clusters has implications not only for overheating of white dwarfs, but also for a number of other signatures involving compact stars. 
Non-annihilating DM can capture in massive stars, self-gravitate when they turn into compact stars, and collapse into black holes that then accrete the stellar material and destroy the star~\cite{Kouvaris:2010jy}.
DM captured in WDs and NSs may annihilate to long-lived mediators that escape the star and decay to SM states that can be detected~\cite{Cermeno_2018,Leane:2021ihh}. 

In certain models dark matter can trigger Type Ia-like supernovae.
This could occur if DM deposited energy in a small pocket of WD material at a rate higher than the energy diffusion rate, as that would trigger runaway nuclear fusion that unbinds the WD.
Observations of the survival of WDs to this date can then be used to place constraints on this mechanism.
This idea has been investigated in the context of DM in primordial black holes depositing energy in WDs via dynamical friction~\cite{GrahamRajendVarelaPBHWD},
non-annihilating particle DM captured by the WD depositing gravitational potential energy via nucleon scattering as it collapses in the WD~\cite{GrahamRajendVarelaPBHWD,Bramante:2015cua} (though see Ref.~\cite{SteigerwaldProfumo:2022pjo}),
heavier-than-10$^{16}$ GeV DM depositing energy via annihilations, decays or nuclear scattering~\cite{Graham:2018efk}, and energy deposition from rapid Hawking radiation emitted by black holes formed in the interior of WDs via DM collapse~\cite{Acevedo:2019gre,Janish:2019nkk}.

In all these cases, a sufficiently high DM density is required to ensure sufficiently high WD capture/encounter rates.
Just as we urge future authors to refrain from using globular clusters to study WD heating, we urge the same of them in studies of other effects of DM on compact stars.

\section{Discussion and Prospects}
\label{sec:discs}

We have investigated whether recent claims in the literature on constraining dark matter capture and annihilations in white dwarfs in the globular cluster M4/NGC6121 are compatible with a first empirical estimate of the DM content in the system.
We have also commented on other mechanisms of probing DM using compact objects residing in globular clusters.
Using line-of-sight stellar velocity and surface luminosity data, we performed an MCMC likelihood analysis and found no evidence for a DM component in M4/NGC6121. 
This sets only an upper bound on the DM content, that still doesn't negate the validity of WD heating constraints.
However, due to irreducibly large uncertainties in the problem and necessarily weak bounds on the DM density in M4, our broad conclusion is that, the WD heating constraints from globular clusters are unreliable\footnote{Ref.~\cite{Cline:2020mdt} refrains from constraining their DM model using M4/NGC6121 after explicitly stating this reason. We encourage other authors to adhere to this spirit.}.
Perhaps our stance is clarified by comparing to the state of affairs in the Galactic Center.
With current stellar kinematic data, the DM density in both globular clusters and the Galactic Center is ``unknown" (but the important difference is that in globular clusters it is also consistent with zero).
The uncertainty in the Galactic Center density propagates into the well-known inconclusivity about the 3.1 TeV thermal wino, the supersymmetric partner of the $W$ boson. 
Due to astrophysical $J$-factors varying by a factor of $>100$, gamma-ray line searches at the Galactic Center by H.E.S.S. would rule out the 3.1 TeV wino for an NFW profile, yet leave it safe for a largely cored Burkert profile~\cite{Cohen:2013ama}.  

The above inconclusivity is, of course, buried within another inconclusivity.
As mentioned in the Introduction, one pathway to form globular clusters is the collapse of massive molecular gas clouds with highly efficient star formation, supported by observations of the Antennae Galaxy merger.
This mechanism requires no DM for forming self-gravitating $\sim 10^5 M_\odot$-heavy dense stellar structures.
Even if DM is required, such as in the pathway initiated by a DM subhalo that is then tidally stripped, the final DM mass in the cluster could be as low as $10^2 \ M_\odot$, weakening the claims of large DM densities affecting compact objects~\cite{Hooper:2010es}.
Finally, even if statistical evidence of a ``dark" component is found, it may not be distinguished from a population of cold stellar remnants~\cite{EvansStrigari:2021bsh}.

One possible way to improve the DM limits in globular clusters, albeit marginally, is to obtain stellar kinematic data in their innermost regions, which is a question of telescope resolution of dense stellar fields. 
Observations of small dispersion speeds in these regions will lead to tighter bounds on the enclosed DM mass $M_{\rm DM} (r)$ at small $r$, which in turn is a tighter bound on $\rho_s r^3_s$ in these regions (as argued in Section~\ref{sec:limits}). 
Of course, improved measurements of proper motions and velocity dispersions in the outer regions of globular clusters would also be helpful, especially to diagnose a flat rotation curve indicative of a DM halo.   
These above improvements are foreseen with the use of {\em Gaia} Data Release 3~\cite{baumgart:gaia:2021} in conjunction with soon-to-be available data from JWST~\cite{2021jwst.prop.bedin}.
It is beyond our scope to identify member WDs of M4/NGC6121 with the appropriate astrometric cuts, derive the corrected color-magnitude diagram with photometric data, etc., however we strongly urge the better-equipped astronomical community to perform this analysis.

While our study focused on M4/NGC6121 to address previous claims about DM heating WDs, this phenomenon may be investigated in other globular clusters as well.
A number of findings have been reported on the presence or absence of DM in $>20$ globular clusters, which we recount in Table~\ref{tab:gclist} and Appendix~\ref{app:otherglobclusts}, and a number of globular clusters have been reported to contain WDs.
Bringing these two classes of studies under one roof would make for important progress in the hunt for DM, a task we leave for future authors. 
We comment more on this in Appendix~\ref{app:otherglobclusts}.

An interesting direction of inquiry is DM capture in neutron stars belonging to globular clusters. 
So far a total of 280 pulsars have been discovered in 38 globular clusters\footnote{\href{https://www3.mpifr-bonn.mpg.de/staff/pfreire/GCpsr.html}{https://www3.mpifr-bonn.mpg.de/staff/pfreire/GCpsr.html}}, but simulations estimate $\Oc (10^{2-3})$ NSs per globular cluster~\cite{Ye:2019luh}.   
While the entire NS population may not be observable in the near future, faint NSs are expected to be either directly or indirectly discovered through surveys and deep field observations. 
For a DM density of 1000 GeV/cm$^3$, NSs could be typically heated to a maximum temperature $\simeq 1.8 \times 10^4$ K, with their spectral distribution peaking near the visible range. 
However, due to the NSs' small radii and large distances from Earth, the observable spectral flux density of $\sim$ picoJansky is several orders below the threshold of current instruments, e.g., nanoJansky at JWST. 
Searches for gamma-ray fluxes from the decay of long-lived mediators produced in the annihilation of DM within NSs in the globular cluster 47 Tuc have been used to set limits on certain DM models~\cite{Leane:2021ihh}, assuming a DM density of 1000 GeV/cm$^3$ within the inner 4 pc and a population of 300-4000 NSs. 
The presence of DM in 47 Tuc is disputed (see Table~\ref{tab:gclist}), with an upper limit on its densities most recently set by Ref.~\cite{Reynoso-Cordova:2022ojo}. 
Assuming an NFW profile, this limit at $r =$ 4 pc ranges between 200$-$5000 GeV/cm$^3$, neither confirming nor denying the assumption in Ref.~\cite{Leane:2021ihh}.

There remains one other spectacular way to constrain DM densities in globular clusters.
And that is a future discovery of DM in underground direct searches.
The cross section and mass of particle DM would inform how efficiently WDs capture it, and thus observations of sufficiently cold WDs in globulars can be used to estimate an upper limit on the ambient DM density.
This method was applied to constrain the DM content of NGC 6397 to smaller than $10^{-3}$ the stellar mass~\cite{Hurst:2014uda} by taking at face value hints seen at the DM experiments CRESST, DAMA, CDMS-Si, and CoGeNT.

\section*{Acknowledgments}

For helpful conversations we thank 
Tom Abel, 
Susmita Adhikari,
Joe Bramante,
and
David Morrissey.
R.G. is supported by MIUR grant PRIN 2017FMJFMW  and 2017L5W2PT.
N.R. thanks the International Centre for Theoretical Sciences for hospitality, where part of this work was completed during the workshop ``Less Travelled Path to the Dark Universe" (code: ICTS/ltpdu2023/3).
J.R-C. is supported by MIUR grant PRIN 2017FMDE.

\appendix

\section{Method for setting dark matter density limits in M4/NGC6121}
\label{app:MCMCpipeline}

The free parameters in Eq.~\eqref{eq:freeparams} that are determined by a likelihood analysis using Eq.~\eqref{eq:sigma_los} are defined as follows.

The stellar density profile is taken as a combination of three Plummer spheres:
\begin{equation}
    \rho_\star (r) = \sum_{j=1}^{3} \frac{3 M_j}{4 \pi a_j^3}\left( 1 + \frac{r^2}{a_j^2} \right)^{-5/2},
\label{eq:stellar_density}
\end{equation}
where $M_j$ and $a_j$ are free parameters.
This description is analogous to a Gaussian decomposition assumed in, e.g., Ref.~\cite{EvansStrigari:2021bsh}. 
The projected surface density can be immediately obtained from the above as
\begin{equation}
    \Sigma_*(r)= \sum_{j=1}^3 \frac{M_j}{\pi a_j^2}\left(1 + \frac{r^2}{a_j^2} \right)^{-2}.
\label{eq:surface_density}
\end{equation}

The parametrisation we use for the velocity anisotropy is:
\begin{equation}
    \beta (r) = \beta_0 + \left( \beta_\infty - \beta_0\right)\frac{1}{1 + (r/r_a)^\eta},
\end{equation}
where $\beta_0$ describes an ``inner" anisotropy, 
$\beta_\infty$ an ``outer" anisotropy, with $r_a$ and $\eta$ respectively the radius and the sharpness of the transition. 

We perform our MCMC analysis using the non-parametric code {\sc gravsphere} \cite{2017MNRAS.471.4541R} with the Python wrapper {\sc pygravsphere} \cite{2020MNRAS.498..144G}. 
Although $M_j$ and $a_j$ will be allowed to vary during the scan, {\sc pygravsphere} first computes the values through an optimization technique and then allows the to vary within the MCMC by a factor of 50\% of the best-fit value.
In addition to using surface density and LOS velocity data, {\sc pygravsphere} also uses the so-called virial shape parameters, which can also help breaking the mass-anisotropy degeracy~\cite{2017MNRAS.471.4541R,2020MNRAS.498..144G},
\begin{subequations}
    \begin{equation}
    \begin{split}
        v_{s1} = \frac{5}{2}\int_0^\infty dr \ r \ G M \rho_\star(r)\left[5 - 2\beta(r)\right]\sigma_r^2  \\ 
        = \int_0^\infty dr \ r \ \Sigma_\star (r)\left \langle v_{\rm{LOS}}^4  \right\rangle,
    \end{split}
    \end{equation}
    \begin{equation}
        \begin{split}
            v_{s2}=\frac{4}{35} \int_0^\infty dr \ r^3 G M \rho_\star (r)\left[7-6\beta (r)\right]\sigma_r^2  \\
            = \int_0^\infty dr \ r^3 \Sigma_\star (r) \left \langle v_{\rm{LOS}}^4 \right  \rangle~.
        \end{split}
    \end{equation}
\end{subequations}

Therefore our likelihood function is given by
\begin{equation}
   -2 \ln{ \mathcal{L}} = \chi_{\rm{LOS}}^2 + \chi^2_{\Sigma_*} + \chi^2_{\rm{VSP,1}} + \chi^2_{\rm{VSP,2}}~.
\end{equation}

From the Bayesian approach we found no preference for a DM component, thus estimating credible intervals on the posterior distributions is a prior-dependent computation. 
To derive upper limits, we use a profile likelihood approach with $\rho_s$ and $r_s$ the parameters of interest, and the resulting 95\% C.L. upper limits are displayed in the top left panel of Figure~\ref{fig:densitylimitsM4NFW}.

\section{Obtaining white dwarf luminosities and temperatures}
\label{app:WDlumitemp}

In this appendix we describe a prescription to convert color-magnitude diagrams (CMDs) of stellar data to temperatures and luminosities.
Specifically, we do this for the WDs identified in the CMD in Fig.~11 of Ref.~\cite{Bedin:2009it} depicting HST/ACS data on M4/NGC6121. 
This CMD is displayed in the $m_{606}-m_{775}$ vs $m_{606}$ plane after correcting for reddening and extinction due to dust.   
The color $m_{606}-m_{775}$ can be used to derive the WD effective temperature, and the Vega magnitude $m_{606}$ to derive the WD luminosity.

 The zero-point used in the HST/ACS system, the ``instrumental zero-point," is the magnitude of an object that produces one count per second. 
 Each zero-point refers to a count rate measured in a specific aperture. 
 For point source photometry, the measurement of counts in a large aperture is not possible for faint targets in a crowded field. Therefore, counts are measured in a small aperture, then an aperture correction is applied to transform the result to an ``infinite" aperture. 

As  discussed in detail in Ref.~\cite{Bedin:2004zm}, the raw data is in the form of the number of total number photo-electrons $I_e$ detected in the exposure time $t_{\rm exp}$. 
This must not be confused with the usual flux expressed in ergs/s/cm$^2$. 
Photometric observations are transformed to Vega-mag using~\cite{Bedin:2004zm}
\begin{eqnarray}
    m_{\rm filt} &=& -2.5 \log_{10} I_e + m_0^{\rm filt} -\Delta m^{\rm filt}_{\rm PSF-AP(r)} \nonumber \\
    && -\Delta m^{\rm filt}_{\rm AP(r)-AP(\infty)}~.
\end{eqnarray}
where $m_0^{\rm filt}$ is  the zero-point of the filter, and the last two terms are aperture corrections.  
Ref.~\cite{Bedin:2009it} provides the left-hand side of the above equation with all the correction factors included.

 \begin{figure}[htb!]
    \centering
    \includegraphics[width=0.45\textwidth]{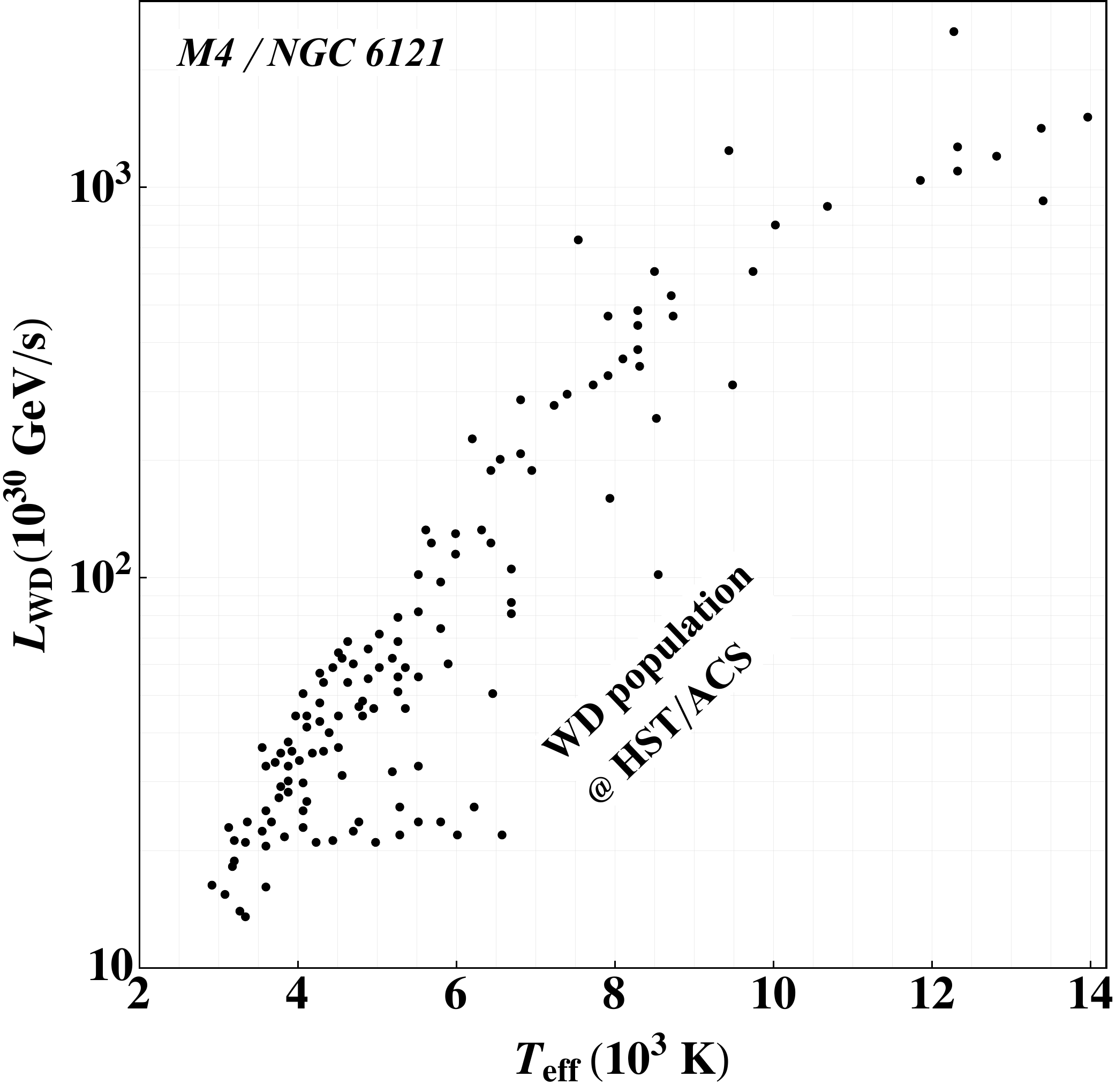}
    \caption{The white dwarf population of M4/NGC6121 observed in HST/ACS in the luminosity-surface temperature plane obtained from the color-magnitude diagram in Ref.~\cite{Bedin:2009it} via the routine described in Appendix~\ref{app:WDlumitemp}.}
    \label{fig:LvTscatter}
\end{figure}

 The detector count rate in a given filter is  
\begin{eqnarray}
    I^{\rm filt}_e &=& A_{\rm tel}\bigg(\frac{R_{\rm WD}}{d}\bigg)^2 \times  \\
    &&10^{-0.4 A_{\rm filt}} \int_{\nu_1}^{\nu_2} d\nu (h\nu)^{-1} B_\nu (\nu, T_{\rm WD}) \epsilon_{\rm filt}(\nu)~, \nonumber
\end{eqnarray}
where $A_{\rm tel}$ is the effective collecting area of the telescope, 
$R_{\rm WD}$ the radius of the WD, 
$d$ its distance from Earth, 
$B_\nu$ the blackbody spectral density, 
$\epsilon_{\rm filt}$ the throughput of the filter, 
and $A_{\rm filt}$ (in magnitude units) accounts for extinction and reddening by intervening dust. 
To obtain the temperature we will always take differences in magnitude, hence the factor $A_{\rm tel}(R_{\rm WD}/d)^2$ will disappear in practice.  
We numerically obtain the color as
\begin{equation}
m_{606} - m_{775} =-2.5 \log_{10}\bigg(\frac{I^{606}_e}{I^{775}_e}\bigg) + m_0^{606} - m_0^{775} + {\rm apr}~,    
\end{equation}
where the last term is the difference in aperture corrections, which turns out to be at the sub-percent level. 

Magnitudes in Ref.~\cite{Bedin:2009it} are reported in the Vega magnitude system, hence 
the WD luminosities can be directly obtained from 
\begin{equation}
m_{\rm WD-Veg}  = -2.5 \log_{10}\bigg(\frac{L_{\rm WD}/d^2_{\rm M4}}{L_{\rm Veg}/d^2_{\rm Veg}} \bigg)|_{606}~.  
\end{equation}
We take the distances $\{d_{\rm M4}, d_{\rm Veg}\}$ = $\{1850, 7.68\}$~pc.
Vega's temperature $T_{\rm Veg} = 9550$~K and radius $R_{\rm Veg} = 2.52 R_\odot$ gives us the blackbody bolometric luminosity $L^{\rm BB}_{\rm Veg}$ via the Stefan-Boltzmann law.
The actual luminosity $L_{\rm Veg} = 37 L_\odot$ gives us the emissivity $L_{\rm Veg}/L^{\rm BB}_{\rm Veg}$, which we use in the above equation.

In practice we use the package {\sc pysynphot}~\cite{pysynphot2013} for the conversion from color vs magnitude to luminosity vs effective temperature.
The result of our conversion is shown in Fig.~\ref{fig:LvTscatter}, which is  slightly different from that in Ref.~\cite{McCullough:2010ai} likely due to the routines in {\sc pysynphot}.

Once $L_{\rm WD}$ and $T_{\rm WD}$ are obtained, the WD radius $R_{\rm WD}$ is obtainable from the blackbody luminosity $L_{\rm WD} = 4\pi \sigma_{\rm SB} R^2_{\rm WD} T^4_{\rm eff}$, where $\sigma_{\rm SB}$ is the Stefan-Boltzmann constant.
We then obtain the WD mass $M_{\rm WD}$ from a mass-radius relation derived by solving the Tolman–Oppenheimer–Volkoff equations for the relativistic Feynman-Metropolis-Teller equation of state (EoS) that models the WD as an isothermal relativistic Fermi gas including Coulomb interactions. We have assumed that the WD is composed entirely of $^{12}$C. 
We obtain a maximum mass of 1.385 $M_\odot$, corresponding to $R_{\rm WD} = 2\times 10^{-3} R_\odot$, in agreement with Ref.~\cite{Rotondo:2011zz}. 
Other EoSs such as the Hamada-Salpeter~\cite{HamadaSalpeter:1961} and Chandrasekhar-Emden EoS predict a critical maximum mass $\approx 1.4\,M_\odot$ for the same composition. 

\section{Dark matter in other globular clusters}
\label{app:otherglobclusts}

There are about 180 globular clusters discovered while their DM content has been searched for in only about 20 of them; see Table~\ref{tab:gclist}.
In this appendix we non-exhaustively review the literature on searches for dark matter in globular clusters and outline some directions for progress.
Numerous techniques have been tried and wide-ranging results have been reported; the goal of this note is to urge the astrophysics community to unify their approaches so that clearer conclusions may be drawn on the important question of the presence of DM in globular clusters.

{\bf NGC 2419.}

The initial population of dark matter in globular clusters can be depleted via dynamical friction of stars ejecting the DM and via tidal stripping by the host galaxy. 
For these reasons, the globular cluster NGC 2419 was selected by Ref.~\cite{Baumgardt2009} to look for DM: its timescales for dynamical friction and relaxation exceed a Hubble time, and its remote location (with a Galactocentric distance $\sim$ 90 kpc) and large mass minimize tidal stripping.
Using radial velocity data from Keck I and an $N$-body fit, these authors find no evidence for DM, supported by their finding that the mass-to-light ratio does not rise toward the outer regions of NGC 2419.
Assuming an NFW profile for DM, they set a 2$\sigma$ limit on the DM mass of $M_{\rm DM} < 10^7 M_\odot$ inside $r =$ 500 pc, equivalently $M_{\rm DM} (r< 260 \ {\rm pc}) \lsim 4 \times 10^6 M_\odot$, corresponding to a limit on the density of 0.7 GeV/cm$^3$.
Ref.~\cite{Conroy2011} also set upper limits on an NFW DM profile, obtaining the very tight $M_{\rm DM}(r < \ {\rm kpc}) < 10^6 M_\odot$, equivalently $M_{\rm DM}(r < 260 \ {\rm pc}) < 2.4 \times 10^4 M_\odot$.

In contrast to these studies, Ref.~\cite{Ibata2012} came to an interesting conclusion. 
When the authors tried to fit stellar kinematic data with a Michie model of stellar distribution and a generalized NFW profile for DM, they found no evidence for DM and set a 99\% C.L. limit of $M_{\rm DM}(r < 260 \ {\rm pc}) < 7.2 \times 10^5 M_\odot$.
However, when they performed a spherical Jeans analysis similar to our work but assuming {\em no} analytic form for the stellar and DM distributions, instead floating 389 free parameters in the solution, they {\em did} find a DM component within 260 pc of mass $\simeq 10^6 M_\odot$, about twice the mass of the stellar component.
This highlights the extreme sensitivity of studies looking for DM in globular clusters to priors and parameterizations, and suggests that almost any conclusion derived from these statistical fits must be taken with a grain of salt.

 $\mathbf{\omega}$ {\bf Centauri/ NGC 5139.}

It is thought that the largest globular cluster observed, the 4 $\times 10^6 M_\odot$-heavy $\omega$Cen, is the tidally stripped relic not of a DM subhalo but of a dwarf galaxy captured by the Milky Way.
Ref.~\cite{Brown:2019whs} applied a spherical Jeans analysis to MUSE and Keck stellar LOS velocity data and Gaia and HST proper motion data, and found evidence for a $\sim 6 \times 10^{5-6} \ M_\odot$ DM component within a 7 pc half-light radius when fitting to an NFW DM profile.
This result was confirmed in Ref.~\cite{EvansStrigari:2021bsh} using updated data from the same sources\footnote{In these studies the total DM mass was fitted simultaneously with the {\em stellar} (as opposed to total) mass-to-light ratio, which were found to be unsurprisingly anti-correlated. 
It may be seen from the posterior distributions of Refs.~\cite{Brown:2019whs,EvansStrigari:2021bsh} that the total mass-to-light ratio is indeed roughly constant across the favored DM mass range.}; it was also argued that this invisible mass component is consistent with a population of stellar remnants.

On the other hand, Ref.~\cite{Reynoso-Cordova:2022ojo} (with one of us as an author) found no evidence for an NFW component of DM in $\omega$Cen, setting instead an upper limit of $M_{\rm DM}(r < 7 \ {\rm pc}) < {\rm few} \ \times 10^5 M_\odot$.
The main difference between these studies is that the latter used LOS dispersion data from MUSE, whereas the former additionally used proper motion data.
Other differences include the modelling of stellar distributions as a sum of Gaussians in the former (the so-called CJAM model) versus a sum of Plummer spheres in the latter, the simpler parametrization of stellar anisotropy in the former, and the log likelihood analysis performed with Bayesian Multinest sampling algorithm in the former versus MCMC in the latter.

{\bf \em Other globular clusters.}

The LOS-dispersion-spherical-Jeans-MCMC analysis of Ref.~\cite{Reynoso-Cordova:2022ojo} set upper limits on NFW DM component for a number of other globular clusters.
These include:

(a) {\bf M22/NGC6656} and {\bf M30/NGC7099}, corroborated by Ref.~\cite{Lane0908}, which looked for a flattening of dispersion profiles at large radii and for large mass-to-light ratios using radial velocity data from Anglo-Australian Telescope's AAOmega spectrograph.
The latter also found no evidence for DM in {\bf M53/NGC5024} and {\bf M68/NGC4590}.

(b) {\bf 47 Tuc/NGC104}, corroborated by Ref.~\cite{Lane0910} using AAOmega.
The latter also found no evidence for DM in {\bf M55/NGC6809}, {\bf NGC 121} and {\bf Kron 3}.
We mention that Ref.~\cite{Brown:2018pwq} explains the observed $\gamma$-ray flux from 47 Tuc/NGC104 using an annihilating DM component.

(c) {\bf NGC 1851, NGC 2808,  NGC 3201, M80/NGC6093, NGC 6752, M2/NGC7089}.

A number of other globular clusters have been studied for the presence of DM, and an array of conclusions drawn from astrophysical arguments.
We refer the reader to Table~\ref{tab:gclist} for a list of references.

{\bf \em White dwarfs in them?}

Luminosity measurements of WDs in globular clusters would be greatly relevant to limiting DM-induced heating if clear evidence for DM content comes up in these systems.
In Table~\ref{tab:gclist} we list references on observations of WDs in various globulars.
The possibility of observing DM-induced WD heating in $\omega$Cen is discussed in Refs.~\cite{Amaro_Seoane_2016,Krall_2018}.
As mentioned in the Discussion, WD heating can be used to {\em limit} DM densities in NGC 6397 if an unambiguous DM signal is found in direct detection experiments~\cite{Hurst:2014uda}; this reasoning of course applies to any globular cluster including the focus of our study, M4/NGC6121.

\bibliography{refs}

\end{document}